\documentclass[prl,twocolumn,showpacs,preprintnumbers,amsmath,amssymb,superscriptaddress]{revtex4-1}

\usepackage{graphicx}
\usepackage{dcolumn}
\usepackage{bm}
\usepackage{epsfig}
\usepackage{amsmath}
\usepackage{amssymb}
\usepackage{color}
\usepackage{bbm}
\usepackage[caption=false]{subfig}
\usepackage{placeins}
\usepackage{soul}

\usepackage[colorlinks=true,citecolor=blue,linkcolor=blue,urlcolor=blue]{hyperref}
\usepackage{cancel}

\def\equationautorefname~#1\null{%
	Eq.~#1\null
}
\def\figureautorefname~#1\null{%
	Fig.~#1\null
}

\begin{document}

\title{Nonreciprocal Bundle Emissions of Quantum Entangled Pairs}
\author{Qian Bin}
\affiliation{School of Physics and Institute for Quantum Science and Engineering, Huazhong University of Science and Technology,
and Wuhan Institute of Quantum Technology, Wuhan 430074, China}

\author{Hui Jing}
\affiliation{Key Laboratory of Low-Dimensional Quantum Structures and Quantum Control of Ministry of Education, Department of Physics and Synergetic Innovation Center for Quantum Effects and Applications, Hunan Normal University, Changsha 410081, China}

\author{Ying Wu}
\affiliation{School of Physics and Institute for Quantum Science and Engineering, Huazhong University of Science and Technology,
and Wuhan Institute of Quantum Technology, Wuhan 430074, China}

\author{Franco Nori}
\affiliation{Theoretical Quantum Physics Laboratory, Cluster for Pioneering Research, RIKEN, Wakoshi, Saitama 351-0198, Japan}
\affiliation{Center for Quantum Computing, RIKEN, Wakoshi, Saitama 351-0198, Japan}
\affiliation{Physics Department, The University of Michigan, Ann Arbor, Michigan 48109-1040, USA}

\author{Xin-You L\"{u}}
\email{xinyoulu@hust.edu.cn}
\affiliation{School of Physics and Institute for Quantum Science and Engineering, Huazhong University of Science and Technology,
and Wuhan Institute of Quantum Technology, Wuhan 430074, China}

\date{\today}

\begin{abstract}

Realizing precise control over multiquanta emission is crucial for quantum information processing, especially when integrated with advanced techniques of manipulating quantum states. Here, by spinning the resonator to induce the Sagnac effect, we can obtain nonreciprocal photon-phonon and photon-magnon super-Rabi oscillations under conditions of optically driving resonance transitions. Opening dissipative channels for such super-Rabi oscillations enables the realization of directional bundle emissions of entangled photon-phonon pairs and photon-magnon pairs by transferring pure multiquanta state to bundled multiquanta outside of the system. This nonreciprocal emission is a flexible switch that can be controlled with precision, and simultaneous emissions of different entangled pairs (such as photon-phonon or photon-magnon pairs) can even emerge but in opposite directions by driving the resonator from different directions. This ability to flexibly manipulate the system allows us to achieve directional entangled multiquanta emitters, and has also potential applications for building hybrid quantum networks and on-chip quantum communications.

 \end{abstract}
\maketitle

Nonreciprocity plays a key role in various devices such as isolators and circulators\,\cite{Potton2004,Lodahl2017MS,Sounas2017Alu}.  Nonreciprocal devices, allowing the flow of signal from one side but not the other, have important applications for the realizations of  noise-free information processing and amplifiers\,\cite{Kamal2011CD,Liu2021ZZ,Sliwa2015HN,Kerckhoff2015LC,Metelmann2015Clerk,Malz2018TB,Shi2015YF}, one-way communications\,\cite{Feng2011AH,Scheucher2016HW}, unidirectional invisibility\,\cite{Lin2011RE,Wang2019RY}, and directional lasing\,\cite{Bahari2017NV,Jiang2018MC,Xu2021LL,Xu2021Song}.  Many nonreciprocal devices have been achieved based on magnetic biasing\,\cite{Saleh2007Teich, Jalas2013PE}, nonlinear optics\,\cite{Fan2012WV,Cao2017WD,Graf2022RS,Yang2019XH}, atomic gases\,\cite{Wang2013ZG,Ramezani2018JW}, optomechanical resonators\,\cite{Peterson2017LC,Bernier2017TK,Shen2016ZC,Lepinay2019DOK,Lepinay2020OKM,Manipatruni2009RL,Shen2023ZC}, spinning resonators\,\cite{Maayani2018DK}, non-Hermitian optics\,\cite{Bender2013FB,Peng2014OL,Chang2014JH}, and acoustic materials\,\cite{Fleury2014SS,Zangeneh-Nejad2018Fleury,Wiederhold2019SA,Coulais2017SA,Cui2018YC}. However,  previous works have mainly focused on classical regimes rather than quantum regimes.  Recently, many nonreciprocal quantum phenomena have been predicted, including single-photon insolation and circulation\,\cite{Tang2022NT,Xia2018NX,Shen2011BS,Dong2021XZ,Ren2022WF}, one-way flow of thermal noise\,\cite{Barzanjeh2018AX}, and nonreciprocal quantum blockade\,\cite{Wang2022XX,Huang2018ML,Li2019HX,Jing2021SX,Shen2020WW,Wang2019WY} and quantum entanglement\,\cite{Ren2022,Jiao2022LL,Jiao2022ZZ}.

Multiquanta physics is an increasing popular research line in the field of quanta state manipulation, with important applications in many fields such as quantum metrology and lithography\,\cite{Giovannetti2006LM,DAngelo2001CY}, biodetection\,\cite{Denk1990SW,Horton2013WK,Chu2011HZ}, and lasing engineering\,\cite{Gauthier1992WM}. Current methods generating multiquanta states mainly rely on post-selection\,\cite{Pan2012CL}, Rydberg atoms\,\cite{ Bienias2014CF,Maghrebi2015GB,Jachymski2016BB,Firstenberg2016AH,Liang2018VC}, and waveguide/cavity/circuit quantum electrodynamics (QED) systems\,\cite{GarciaMaraver2004CE,Koshino2013IY, Douglas2016CC,Sanchez-Burillo2016MMGR,Gonzalez-Tudela2017PK,Ota2011IK,Munoz2014VT, Sanchez-Munoz2015LT, Hargart2016MRC, Chang2016GTM, Munoz2018LV,Satzinger2018ZC,Qin2019MM,Bin2020LL,Cosacchi2022MPS,Zou2022LL,Gou2022HW,Bin2021WL,Ma2021LR, Liu2023HT,Dong2019Li, Ren2021DX}. In particular, the continuous generation of multiquanta state has been proposed under atom-driven systems\,\cite{Munoz2014VT,Sanchez-Munoz2015LT,Munoz2018LV,Bin2020LL,Zou2022LL,Gou2022HW,Bin2021WL,Ma2021LR, Liu2023HT,Deng2021SY,Yuan2023XD}, where the unit of emission is replaced by a bundle of strongly correlated $n$ quanta. This multiquanta bundle emission means that the quantum emitter releases energy in the group of $n$ quanta, which is different from the lasing\,\cite{Bahari2017NV,Jiang2018MC,Xu2021LL,Xu2021Song} that releases energy in the form of single-quanta emission and investigates the amplification of phonons or magnons. This bundle emission is a powerful source for achieving Heisenberg-limited quantum metrology\,\cite{Deng2023LC,Chu2018KY}. However, as far, the realization of bundle emission has not reach exquisite controllability such as strong nonreciprocity. Previously suggested bundle emissions were all reciprocal, and overcoming this limitation to achieve controllable directional bundle emission has remained challenging. Exquisite controllability for bundle emission is highly desirable and may have an important role in the design of directional multiquanta emitters.

Here, we propose an experimentally feasible scheme to achieve switchable nonreciprocal  bundle emissions of quantum entangled pairs in a setup consisting of a strongly driven atom coupled to an optically pumped opto-magnetic resonator. Spinning the resonator introduces a frequency split in the countercirculating modes via the Sagnac effect\,\cite{Maayani2018DK}. The Jaynes-Cummings (JC) interaction combined with linear interaction in the resonator can jointly induce the resonant excitation of the dressed atom together with a single photon and a single phonon (or magnon), when the resonator is driven in a chosen direction. The combination of resonator spinning and optically driven resonance transition leads to photon-phonon (or photon-magnon) super-Rabi oscillations\,\cite{Strekalov2014} that occur in one direction but not in the other. Combined with system dissipations, this super-Rabi oscillation can transfer the pure multiquanta state to bundles of two strongly correlated quanta of completely different natures outside of the system, achieving nonreciprocal bundle emission of quantum pairs. 

\begin{figure}
\includegraphics[width=8.7cm]{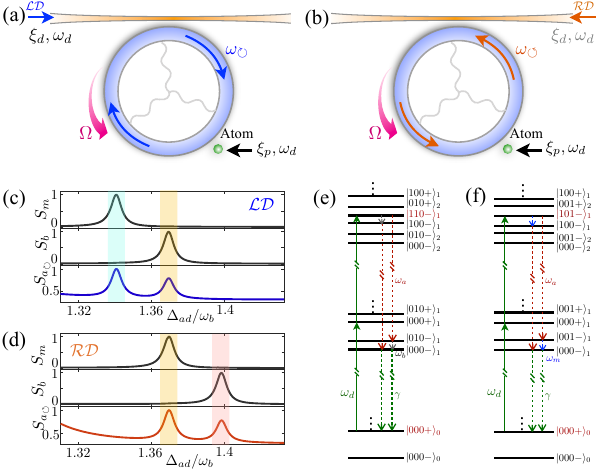}\\
\caption{ (a,b) Schematic of the model.  Fixing the resonator spinning along the counterclockwise direction, the photon-phonon (photon-magnon) pairs arise by driving the resonator from the (a) left  ($\mathcal{LD}$) or (b) right ($\mathcal{RD}$). (c-d) Excitation spectra  $S_o(\epsilon)$ [$o=a,b,m$ correspond to photon, phonon, and magnon modes] versus $\Delta_{ad}/\epsilon$ for (c) $\mathcal{LD}$ and  (d) $\mathcal{RD}$, where $\Delta_{\sigma a}/\omega_b=-3.1$, $\omega_{m}/\omega_b=1.05$, $|\Delta_{F}|/\omega_b=0.025$,  $\lambda_{a\sigma}/\omega_b=0.3$,  $\lambda_{ab}/\omega_b=\lambda_{am}/\omega_b=0.022$, $\xi/\omega_b=0.8$, $\gamma/\omega_b=0.001$, and $\kappa/\omega_b=0.005$. (e) Photon-phonon and (f) photon-magnon transitions in the Mollow ladder from $|+\rangle$ to $|-\rangle$, where the states $|000-\rangle_2$ and $|110\rangle_1$ ($|101+\rangle_1$) are degenerate in (e) [(f)]. After a subsequent emission, the system goes back to the state $|000+\rangle_0$.}\label{fig1}
\end{figure}

Accompanied by this emission, the hybrid system also enables the occurrence of photon-phonon (or photon-magnon) bipartite entanglement and photon-phonon-atom (or photon-magnon-atom) tripartite entanglement. Thus, our proposal continuously generates nonreciprocal entangled photon-phonon (or photon-magnon) pairs. Additionally, entangled photon-phonon and photon-magnon pairs can be emitted simultaneously in opposite directions when the resonator is driven from different directions. These directional hybrid quantum effects, unattainable in purely optical or reciprocal systems, offer new resources for one-way quantum communications, such as directionally transferring quantum information with quantum entangled pairs in future quantum networks\,\cite{Kimble2008, Dong2015WW, Riedinger2016, Wehner2018EH}. This unidirectional  flow can prevent backscattering in quantum processing, as transmission signals are strictly directed along a predefined pathway\,\cite{Wang2009CJ, Chen2011HD}. It also offers potential application for nonreciprocal quantum metrology, e.g., utilizing nonreciprocality  to eliminate the interference typically caused by back-reflections, and applying quantum entangled states to enhance quantum sensing, overcoming the standard quantum limit and approaching the Heisenberg limit\,\cite{Xia2023AP,Gessner2018PS, Duarte2021Taylor}.

\emph{Model and hybrid super-Rabi oscillations}.---We consider theoretically a hybrid system, with a  coherently driven two-level atom coupled to a spinning yttrium iron garnet (YIG) microresonator with coupling strength $\lambda_{a\sigma}$, as shown in Fig.\,\ref{fig1}(a).  The optical modes in the microcavity can be modulated by the mechanical breathing mode\,\cite{Aspelmeyer2014KM} and magnetization procession\,\cite{Chai2022SZ}, when an optical pump with frequency $\omega_d$ drives at the optical modes.  The resonator spins  at an angular velocity $\Omega$, causing the resonance frequencies of the countercirculating optical modes to undergo an opposite Fizeau shift, i.e., $\omega_a\to\omega_a+\Delta_F$, with $\Delta_F=\pm r n_r \omega_a\Omega[1-1/n_r^2-(\lambda/n_r)(d n_r/d\lambda)]/c$. Here $\omega_a$ is the frequency of the optical modes of a stationary resonator, $r$ is the resonator radius, $n_r$ is the refractive index of the material, and $\lambda$ ($c$) is the wavelength (speed) of light in vacuum.  The dispersion term $(\lambda/n_r)(d n_r/d\lambda)$ characterizes the relativistic origin of the Sagnac effect and is typically small, and $\Delta_F>0$  and $\Delta_F<0$ correspond to driving the resonator from the left and right, respectively.  In the frame rotating at $\omega_d$, the system effective Hamiltonian derived by a standard linearization procedure is ($\hbar=1$)\,\cite{supp}
\begin{align}\label{eqm01}
H\!=&(\Delta_{ad}\!+\!\Delta_F) a^\dag a+\omega_b  b^\dag b+\omega_m m^\dag m+\Delta_{\sigma d} \sigma^\dag \sigma\nonumber\\
&+\lambda_{ab}(a^\dag+ a)(b^\dag+b)+\lambda_{am}(a^\dag + a)(m^\dag+m)\nonumber\\
&+\lambda_{a\sigma} (a \sigma^\dag+a^\dag \sigma)+\xi(\sigma^\dag+ \sigma),
 \end{align}
where $a$ ($b$, $m$) is the annihilation operator of the optical (mechanical, magnon) mode, $\sigma=|g\rangle\langle e|$ is the lowering operator of the atom, $\Delta_{ad}$ ($\Delta_{\sigma d}$) is the laser detuning with respect to the optical mode (atom), with $\Delta_{\sigma a}=\Delta_{\sigma d}-\Delta_{ad}$. Moreover, $\omega_b$ ($\omega_m$) is the frequency of the mechanical (magnon) mode, $\lambda_{ab}$ ($\lambda_{am}$) is the effective linear coupling strength that is proportional to the classical cavity amplitude, and $\xi$ is the effective driving strength. Strong driving on the atom can dress the atomic levels, forming a Mollow ladder of manifolds. As shown in Figs.\,\ref{fig1}(e,f), each manifold includes many dressed states $|n_a n_b n_m\pm\rangle$, where $n_a (n_b, n_m)$ represents the photon (phonon, magnon) number.  This is different from the well-known Mollow ladder in the single atom and cavity QED systems\,\cite{Mollow1969, Cohen-Tannoudji1977Reynaud, Zakrzewski1991LM, Ulhaq2012We,Carreno2017VL, Nienhuis1993,Valle2012GTL, Munoz2014VT,Sanchez-Munoz2015LT,Munoz2018LV,Bin2020LL}. The dressed states of the atom  $|\pm\rangle = c_{\pm}|g\rangle\pm c_{\mp}|e\rangle$ have corresponding  eigenvalues $E_{|\pm\rangle}=\Delta_{\sigma d}/2\pm\sqrt{\Delta_{\sigma d}^2+4\xi^2}/2$, where $c_{\pm}=\sqrt{2}\xi/[\Delta_{\sigma d}^2+4\xi^2\pm\Delta_{\sigma d}\sqrt{\Delta_{\sigma d}^2+4\xi^2}]^{1/2}$\,\cite{supp}.

 First, when  the spinning resonator is driven from the left, and the total energy of  a single clockwise photon and a single magnon matches with the transition between the $|+\rangle$ and $|-\rangle$, i.e.,  $(\Delta_{ad}+|\Delta_F|)+ \omega_m-\sqrt{\Delta_{\sigma d}^2+4\xi^2}\approx 0$\,\cite{supp}, the photon-magnon resonance transition from the state $|000+\rangle$ to the photon-magnon state $|101-\rangle$ can be induced by combining the JC interaction and linear optomagnetic interaction. This is verified by the super-Rabi oscillation $|000+\rangle\leftrightarrow|101-\rangle$ shown in Fig.\,\ref{fig2}(a). In contrast,  when the spinning resonator is driven from the right, there is no transition to multiquata states due to detuning\,\cite{supp}, as shown in the inset of  Fig.\,\ref{fig2}(a). This nonreciprocal generation of photon-magnon states is also shown by the excitation spectra in Figs.\,\ref{fig1}(c,d).  The blue area in Fig.\,\ref{fig1}(c) corresponds to the excitation of the photon-magnon state, but this excitation in the same parameter regime cannot be seen in Fig.\,\ref{fig1}(d). Second, by varying the driving frequency detuning at the pink area of Fig.\,\ref{fig1}(d), the transition from $|000+\rangle$ to $|110-\rangle$ can be induced by the combination of JC interaction and  linear optomechnical interaction.  Figure\,\ref{fig2}(b) shows that photon-phonon super-Rabi oscillations occur when driving the resonator from the right but not from the left.  Lastly, when the detuning is varied at the orange areas of Figs.\,\ref{fig1}(c,d),  the photon-phonon resonance transition can be induced when  the resonator is driven from the left, while the photon-magnon resonance transition can also be induced when the resonator is driven from the right\,\cite{supp}. Figure\,\ref{fig2}(c) shows the nonreciprocal photon-phonon and photon-magnon super-Rabi oscillations.  Note that the occurrence of super-Rabi oscillations is a crucial step for multiquanta emissions. 

\begin{figure}
\includegraphics[width=8.7cm]{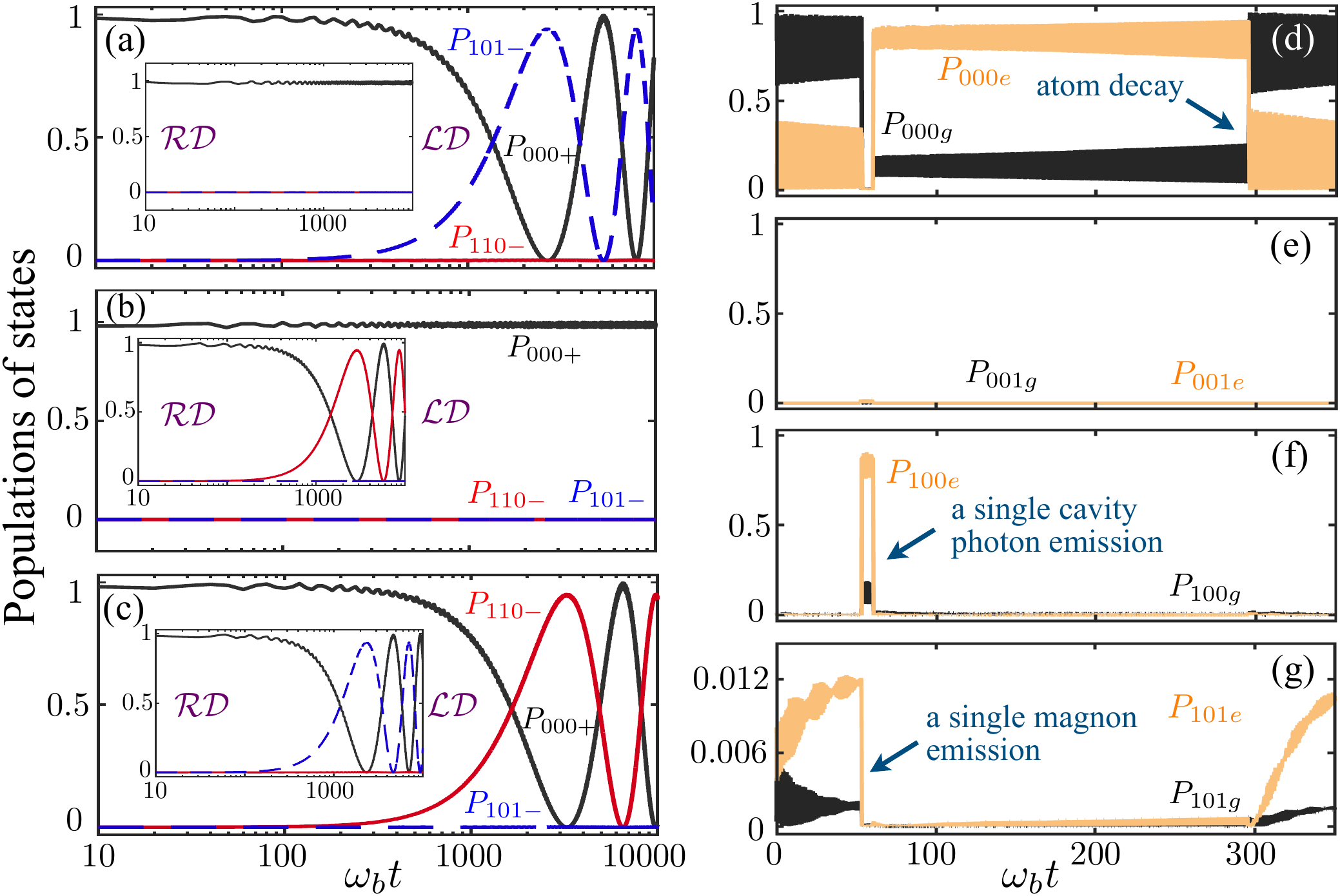}\\
\caption{ (a--c) The population dynamics of the system state  at the photon-phonon (photon-magnon) bundle resonances. (d--g) A quantum trajectory during photon-magnon emission in the same regime as in (a), but in the presence of dissipation. Panels (a), (b), and (c) correspond to the blue, pink, and orange areas in Figs.\,\ref{fig1}(c,d), and $P_{n_a n_b n_m l}=|\langle n_a n_b n_m l|\psi(t) \rangle|^2$ ($n_a,n_b,n_m=0,1$ and $l=+,-,e/g$).}\label{fig2}
\end{figure}

\begin{figure}
\includegraphics[width=8.7cm]{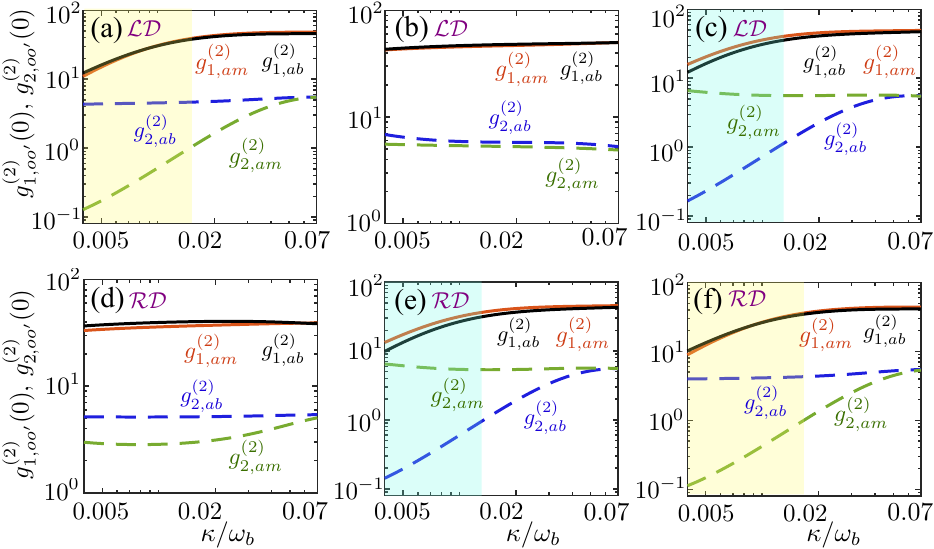}\\
\caption{ Cross correlation  $g_{1, oo'}^{(2)}(0)$  and bundle correlation $g_{2, oo'}^{(2)}(0)$ versus $\kappa/\omega_b$ when  driving the resonator from the (a-c) left and (d--f) right. Panels (a,d), (b,e), and (f,g) correspond to the resonance regimes indicated in blue,  pink, and orange areas of Figs.\,\ref{fig1}(c,d). The shaded areas correspond to the regime of  $g_{1,ab}^{(2)}(0)>1$ ($g_{1,am}^{(2)}(0)>1$) and $g_{2,ab}^{(2)}(0)<1$ ($g_{2,am}^{(2)}(0)<1$).  
Other system parameters are the same as in Figs.\,\ref{fig1}(c,d). 
}\label{fig3}
\end{figure}

\emph{Nonreciprocal emission of entangled photon-phonon and photon-magnon pairs}.---System dissipations need to be considered here for triggering quantum emission. The dynamical behavior of the dissipative system can be described by a quantum master equation 
\begin{align}\label{eqm02}
\frac{d\rho}{dt}=&-i[H,\rho]+\sum_{o=a,b,m}\kappa\mathcal{L}[o]\rho+\gamma\mathcal{L}[\sigma]\rho,
 \end{align}
 where $\mathcal{L}[o]\rho=(2o\rho o^\dag-\rho o^\dag o-o^\dag o\rho)/2$ is a Lindblad term, $\gamma$ is the atomic decay rate, and $\kappa$ is the decay rate of photon (phonon, magnon). System dissipations enable the above photon-phonon (photon-magnon) states to be the bundles of strongly correlated single photons and  phonons (magnons) outside of the system. In Figs.\,\ref{fig2}(d-g), we depict a brief quantum trajectory to demonstrate the bundle emissions by calculating the populations of states using a quantum Monte Carlo method.  Initially, the system is mainly in a superposition state of the vacuum state and photon-magnon state. Triggered by dissipation, a quantum collapse is likely to occur into the photon-magnon state, resulting in the emission of a single magnon and leaving the system in the single-photon state. Subsequently, a single cavity photon is emitted within a very short temporal window, completing the photon-magnon bundle emission.   Consequently, the system emits two strongly correlated single photon and single magnon, leaving it in a vacuum state. Note that there is randomness in the emission sequence of the photon and magnon.  Following a direct photon emission from the atom decay, a photon-magnon state is again prepared for the next bundle emission. In the subsequent cycle, the system undergoes the same cascade emission of pair of photon-magnon, each accompanied by a single photon emission that is at another frequency and does not disrupt the bundle. The processes of the photon-phonon and photon-magnon bundle emissions  are similar.

 

 The strong correlation between the single photon and single phonon (magnons) in a bundle can be determined by the correlation $g_{1, oo'}^{(2)}(0)=\langle o^\dag o o'^\dag o' \rangle/(\langle o^\dag o\rangle \langle o'^\dag o'\rangle)$ \,\cite{Glauber1963, Scully and Zubairy1997,Rebic2002PT},  as shown in  Fig.\,\ref{fig3}. However,  the correlations between bundles themselves here need to be quantified by the generalized second-order correlation of the bundle\,\cite{Munoz2014VT}
 \begin{align}\label{eqm03}
g_{2, oo'}^{(2)}(\tau)=\frac{\langle (oo')^{\dag}(0) (oo')^{\dag}(\tau) (oo')(\tau) (oo')(0)\rangle}{\langle (oo')^{\dag}(0) (oo')(0)\rangle \langle (oo')^{\dag}(\tau) (oo')(\tau)\rangle},
 \end{align}
where $\tau$ is time-delay. Figures \ref{fig3}(a,c,e,f) show the evolution of photon-phonon (photon-magnon) bundle from antibunching, through coherence,  to bunching.   However, as shown in the shaded areas in Figs.\,\ref{fig3} (a,d) [Figs.\,\ref{fig3} (b,e)], the photon-magnon (photon-phonon) pairs only occur when the resonator is driven in one direction,  due to the absence of the excitation of multiquanta states for driving in the other; i.e., there is a nonreciprocal generation of antibunched photon-magnon (photon-phonon) pairs. In Figs.\,\ref{fig3} (c,f), we present that in this parameter regime shown by the orange areas of Figs.\,\ref{fig1}(c,d),  the nonreciprocal emissions of antibunched photon-phonon and photon-magnon pairs simultaneously emerge in opposite directions when driving the resonator from different directions.  The  strongly antibunched bundles actually correspond to single photon-phonon  (photon-magnon) pairs with $g_{2,ab}^{(2)}(0)\ll1$ ($g_{2,am}^{(2)}(0)\ll1$), analogous to single photon with $g_{1,aa}^{(2) } (0)\ll1$. Moreover, it is also possible to realize nonreciprocal emissions of entangled photon-phonon and photon-magnon pairs in our proposal, since the system involves multiple degrees of freedom of completely different nature.

To verify the above emitted quantum pairs are entangled, we calculate the entanglement between different particles\,\cite{Amico2008FO,Walter2016GE,Fabre2020Treps,  Coffman2000KW, Adesso2006lluminati, Rebic2010MM} based on quantum Fisher information (QFI) and covariance. For  an arbitrary separable quantum states of $N$ particles, the QFI $F_Q$  and  covariance ${\rm Var} (A_j)_{\rho}$ satisfy\,\cite{Gessner2016PS,Tian2002XS}
 \begin{align}\label{eqm04}
F_Q[\rho_{\rm sep}, \sum_{j=1}^N A_j]\leq 4\sum_{j=1}^N {\rm Var}(A_j)_{\rho_{\rm sep}}\equiv B_n,
 \end{align}
where $A_j$ is a local observable for the $j$th particle.  The violation of Eq.\,(\ref{eqm04}) for any choice of $A_j$ is a sufficient criterion for entanglement. To find the optimal local operator, we write  $A_j={\bf c}_j \cdot {\bf A}_j$, where  ${\bf A}_j=(A_j^{(1)}, A_j^{(2)}, \dots)^T$ and ${\bf c}_j=(c_j^{(1)}, c_j^{(2)}, \dots)$ are the set of operators for the $j$th particle and corresponding coefficients. The full operator is then $A({\bf c})= {\bf c}\cdot{\bf A}$, with the  combined vectors ${\bf c}=({\bf c}_1, {\bf c}_2, \dots, {\bf c}_N)$ and the operators ${\bf A}=[{\bf A}_1, \dots, {\bf A}_N]^T$.  In our model,  we select ${\bf A}_j= (\sigma_j^x, \sigma_j^y, \sigma_j^z)^T $ for the atom and ${\bf A}_j= (x_j, p_j, x_j^2, p_j^2, (x_jp_j+p_jx_j)/2)^T $ for the mode $o$,  where $x=o+o^\dag$   and $p=-i(o-o^\dag)$\,\cite{Tian2002XS}.  The QFI and covariance can be written as $F_Q[\rho, A({\bf c})]={\bf c}Q_{\rho}^{\mathcal{A}} {\bf c}^T$ and $\sum_{j=1}^N {\rm Var}({\bf c}_j \cdot {\bf A}_j)={\bf c}\Gamma_{\rho}^{\mathcal{A}} {\bf c}^T$, respectively\,\cite{Hyllus2010GS}. Here,  $Q_{\rho}^{\mathcal{A}}$ has elements $(Q_{\rho}^{\mathcal{A}})_{jj'}^{mm'}=2\sum_{k,k'}\frac{(p_k-p_{k'})^2}{p_k+p_{k'}}\langle \psi_{k}|A_j^{(m)}|\psi_{k'}\rangle \langle \psi_{k'}|A_{j'}^{(m')}|\psi_{k}\rangle$ using the spectral decomposition $\rho=\sum_k p_k |\psi_k\rangle \langle \psi_k|$\,\cite{Braunstein1994Caves}, and $\Gamma_{\rho}^{\mathcal{A}}$ has elements $(\Gamma_{\rho}^{\mathcal{A}})_{jj'}^{mm'}={\rm Cov} (A_j^{(m)},A_{j'}^{(m')})_{\rho}$ and $(\Gamma_{\rho}^{\mathcal{A}})_{jj}^{mm'}=0$. The optimal   ${\bf c}$ can be found by calculating the optimal eigenvalue and corresponding eigenstate of the matrix $Q_{\rho}^{\mathcal{A}} -\Gamma_{\rho}^{\mathcal{A}}$\,\cite{supp}.

\begin{figure}
\includegraphics[width=7cm]{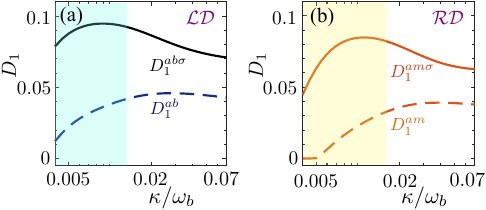}\\
\caption{ (a-b) The Quantities $D_{1}^{ab(m)\sigma}$ and $D_{1}^{ab(m)}$ characterizing the photon-phonon-atom (photon-magnon-atom) entanglement and photon-phonon (photon-magnon) entanglement versus $\kappa/\omega_b$. The shaded areas correspond to the antibundle bundle regime as shown in Figs.\,\ref{fig3}(c,f).  System parameters are the same as in Figs.\,\ref{fig3}(c,f). 
}\label{fig4}
\end{figure}

    A photon-phonon separable quantum state and a photon-phonon-atom fully separable quantum state can be given by $\rho_{ab}=\sum_k p_k\rho_a^k\otimes\rho_b^k$ and $\rho_{ab\sigma}=\sum_{k'} p_{k'}\rho_a^{k'}\otimes\rho_b^{k'}\otimes\rho_{\sigma}^{k'}$ ($\rho_j$ is the reduced density operator), respectively. If the quantity $W^{ab}_1=F_Q[\rho_{ab}, A({\bf c})]-B_1(\rho_{ab})>0$, the photon and phonon modes are entangled. Similarly, if $W^{ab\sigma}_1=F_Q[\rho_{ab\sigma}, A({\bf c})]-B_1(\rho_{ab\sigma})>0$, there is entanglement between the atom, photon mode, and phonon mode\,\cite{supp,Tian2002XS}. Here it is only necessary to replace mode $b$ with mode $m$ to characterize the entanglement between the magnon mode and others. Corresponded to the parameter regime in Figs.\,\ref{fig3} (c,f), in Fig.\ref{fig4} we show the quantities $D_{1}^{ab(m)}={\rm max}\{0, W^{ab(m)}_1\}$ and $D_{1}^{ab(m)\sigma}={\rm max}\{0, W^{ab(m)\sigma}_1\}$ versus  $\kappa/\omega_b$. Comparing Fig.\,\ref{fig3}(c,f) and Fig.\,\ref{fig4}, it is clear that there is the photon-phonon  bipartite entanglement with  $D^{ab}_1>0$ and photon-phonon-atom  tripartite entanglement with $D^{ab\sigma}_1>0$ in the antibunched bundle regime when driving the resonator from the left,  while there is the photon-magnon  bipartite entanglement with  $D^{am}_1>0$ and photon-magnon-atom  tripartite entanglement with $D^{am\sigma}_1>0$ in the antibunched bundle regime when driving the resonator from the right. This confirms that nonreciprocal generation of antibunched photon-phonon (photon-magnon) entangled pairs can be achieved simultaneously  but in opposite directions when we drive the resonator from different directions.  The proposed hybrid system could serve as a nonreciprocal quantum emitter of entangled photon-phonon (photon-magnon) pairs.

\emph{Experimental feasibility and conclusions.}---Our proposal could be implemented in a hybrid setup\,\cite{Aspelmeyer2014KM, Chai2022SZ}, where  a coherently driven two-level atom (e.g., rubidium or cesium atom) is coupled to a spinning YIG microsphere fixed on a rotating platform\,\cite{Maayani2018DK}. The microsphere supports two countercirculating optical whispering gallery modes (WGMs), which are strongly driven  by the input light through a high-index prism\,\cite{Aspelmeyer2014KM}.  The WGMs can be modulated by  the mechanical breathing mode\,\cite{Aspelmeyer2014KM} and magnetization procession\,\cite{Chai2022SZ} when the input light pumps at the WGM.  The microsphere resonator can also be replaced with the toroidal or bottle microresonator\,\cite{Zhang2016ZZ}.  The atom is cooled and trapped near the microresonator using the magneto-optical trap and optical dipole trap\,\cite{OShea2013JV, Shomroni2014RL, Will2021MR,  Aoki2006DW, Alton2011SA, Dayan2008PA, Junge2013OSV,Thompson2013TdL}. The resonator’s evanescent field, which decays exponentially outside its surface, interacts with the atom. This interaction strength depends on the resonator’s properties and the distance of the atom to its surface. Placing the atom closer to the resonator\,\cite{Alton2011SA, Will2021MR}, scaling down the size of the microresonator to reduce the mode volume\,\cite{Buck2003Kimble, Spillane2005KV},  or  employing  other physical methods such as quantum squeezing\,\cite{Lu2015WJ, Qin2018ML, Burd2021SK, Burd2024KS, Qin2024Nori} can increase the coupling strength (more details in the Supplemental Material).   Additionally,  the phonons  (magnons) counts and other types of statistical processing of photons and phonons (magnons)  can be measured by applying an auxiliary system to convert the mechanical (magnon) signals\,\cite{Meenehan2015MM,Bender2019KB}. The multipartite entanglement can be characterized by a class of nonlinear squeezing parameters \,\cite{Tian2002XS,Gessner2019SP}.  With this special design, we theoretically  predict that the nonreciprocal entangled photon-phonon pairs and photon-magnon pairs with corresponding antibunching $g^{(2)}_{2,ab}=0.49$ and $g^{(2)}_{2,am}=0.29$  could be achieved ($\omega_b/2\pi=2\,{\rm GHz}$, $\omega_{m}/2\pi=2.1\,{\rm GHz}$,  $\Delta_{\sigma a}/2\pi=-6.2\,{\rm GHz}$,  $|\Delta_{F}|/2\pi=50\,{\rm MHz}$, $\lambda_{a\sigma}/2\pi=600\,{\rm MHz}$, $\lambda_{ab}/2\pi=\lambda_{am}/2\pi=44\,{\rm MHz}$, $\xi/2\pi=1.6\,{\rm GHz}$, $\gamma/2\pi=2\,{\rm MHz}$, and $\kappa/2\pi=16\,{\rm MHz}$)\,\cite{Aspelmeyer2014KM, Chai2022SZ, Zhang2016ZZ, OShea2013JV, Shomroni2014RL, Will2021MR,  Aoki2006DW, Alton2011SA, Dayan2008PA, Junge2013OSV, Buck2003Kimble, Spillane2005KV, Thompson2013TdL}.

In conclusion, we have shown how to achieve and exquisitely manipulate nonreciprocal bundle emission of entangled photon-phonon pairs and photon-magnon pairs. The system enables the simultaneous emission of entangled photon-phonon and entangled photon-magnon pairs, but in opposite directions,  by driving the resonator from different directions. This work establishes a connection between nonreciprocal physics and multiquanta emission, providing a way to manipulate multiquanta states using nonreciprocal devices in optics\,\cite{Fan2012WV,Cao2017WD,Yang2019XH}, atomic gases\,\cite{Wang2013ZG,Ramezani2018JW}, and acoustic materials\,\cite{Fleury2014SS,Zangeneh-Nejad2018Fleury,Wiederhold2019SA}.  The proposed hybrid system can serve as a directional quantum emitter of entangled pairs, continuously preparing entangled sources.  This nonreciprocal entangled source may provides unconventional resources  for quantum  communication\,\cite{Kimble2008, Dong2015WW, Riedinger2016, Wehner2018EH} and metrology\,\cite{Xia2023AP,Gessner2018PS, Duarte2021Taylor}.

X.L. is supported by  the National Key Research and Development Program of China grant 2021YFA1400700 and the National Natural Science Foundation of China  (Grant No.\,11974125). Q.B. is  supported by the National Natural Science Foundation of China  (Grant No.\,12205109). H.J. is supported by the National Natural Science Foundation of China (Grants No. 11935006 and No. 11774086) and the Science and Technology Innovation Program of Hunan Province (Grant No. 2020RC4047). F.N. is supported in part by: Nippon Telegraph and Telephone Corporation (NTT) Research, the Japan Science and Technology Agency (JST) [via the Quantum Leap Flagship Program (Q-LEAP), and the Moonshot R $\&$ D Grant Number JPMJMS2061], the Asian Office of Aerospace Research and Development (AOARD) (via Grant No. FA2386-20-1-4069), and the office of naval  research (ONR). The computation iscompleted in the HPC Platform of Huazhong University of Science and Technology.

\onecolumngrid
\clearpage
\setcounter{equation}{0}
\setcounter{figure}{0}
\setcounter{table}{0}
\setcounter{page}{6}
\setcounter{section}{0}
\makeatletter
\renewcommand{\theequation}{S\arabic{equation}}
\renewcommand{\thefigure}{S\arabic{figure}}
\renewcommand{\bibnumfmt}[1]{[S#1]}
\renewcommand{\citenumfont}[1]{S#1}
\begin{center}
        \textbf{Supplemental Material for ``Nonreciprocal Bundle Emissions of Quantum Entangled Pairs"}
\end{center}

\title{Supplemental Material for ``Nonreciprocal Bundle Emissions of Quantum Entangled Pairs"}
\date{\today}

\title{Supplemental Material for ``Nonreciprocal Bundle Emissions of Quantum Entangled Pairs"}
\author{Qian Bin}
\affiliation{School of Physics and Institute for Quantum Science and Engineering, Huazhong University of Science and Technology,
and Wuhan Institute of Quantum Technology, Wuhan 430074, China}

\author{Hui Jing}
\affiliation{Key Laboratory of Low-Dimensional Quantum Structures and Quantum Control of Ministry of Education, Department of Physics and Synergetic Innovation Center for Quantum Effects and Applications, Hunan Normal University, Changsha 410081, China}

\author{Ying Wu}
\affiliation{School of Physics and Institute for Quantum Science and Engineering, Huazhong University of Science and Technology,
and Wuhan Institute of Quantum Technology, Wuhan 430074, China}

\author{Franco Nori}
\affiliation{Theoretical Quantum Physics Laboratory, Cluster for Pioneering Research, RIKEN, Wakoshi, Saitama 351-0198, Japan}
\affiliation{Center for Quantum Computing, RIKEN, Wakoshi, Saitama 351-0198, Japan}
\affiliation{Physics Department, The University of Michigan, Ann Arbor, Michigan 48109-1040, USA}

\author{Xin-You L\"{u}}
\email{xinyoulu@hust.edu.cn}
\affiliation{School of Physics and Institute for Quantum Science and Engineering, Huazhong University of Science and Technology,
and Wuhan Institute of Quantum Technology, Wuhan 430074, China}

\date{\today}
\maketitle
In this Supplemental Material, we present the technical details of the effective Hamiltonian and multiquanta resonances discussed in the main text. In Sec.\,I,  we derive the system's effective Hamiltonian using the standard linearization procedure. In Sec.\,II, we provide a detailed derivation of multiquanta resonance conditions within the Mollow regime. In Sec.\,III,  we discuss the entanglement witness based on quantum Fisher information and covariance. In Sec.\,IV, we discuss the experimental feasibility of our theoretical model in detail.

\section{I.\,Derivation of the effective Hamiltonian via the quantum master equation}\label{section1}
The Hamiltonian of the system we consider is
\begin{align}\label{eqI01}
H_t=&(\omega_a+\Delta_F)a^\dag a +\omega_b b^\dag b+ \omega_m m^\dag m+\omega_{\sigma}\sigma^\dag \sigma+\lambda_{ab}'a^\dag a(b^\dag +b)+\lambda_{am}' a^\dag a (m^\dag +m)+\lambda_{a\sigma}'(a^\dag \sigma+a\sigma^\dag)\nonumber\\
&+\xi_d(a^\dag e^{-i\omega_d t}+a e^{i\omega_d t})+\xi_p(\sigma^\dag e^{-i\omega_d t}+\sigma e^{i\omega_d t}),
 \end{align}
where $a$ ($b$, $m$) is the annihilation operator of the optical (mechanical, magnon) mode with corresponding frequency $\omega_a$ ($\omega_b$, $\omega_m$),  $\sigma=|g\rangle\langle e|$ is the lowering operator of the atom with frequency $\omega_{\sigma}$, $\Delta_F$ is the Fizeau shift,  $\lambda_{a\sigma}'$ ($\lambda_{ab}'$, $\lambda_{am}'$) is the coupling strength between the optical mode and atom (mechanical mode, magnon mode), $\xi_d$ ($\xi_p$) and $\omega_d$ are the amplitude and frequency of the driving at the optical mode (atom), respectively.  In the frame rotating with frequency $\omega_d$, Eq.\,(\ref{eqI01}) becomes
\begin{align}\label{eqI02}
H_t=&(\omega_a-\omega_d+\Delta_F)a^\dag a +\omega_b b^\dag b+ \omega_m m^\dag m+(\omega_{\sigma}-\omega_{d})\sigma^\dag \sigma+\lambda_{ab}'a^\dag a(b^\dag +b)+\lambda_{am}' a^\dag a (m^\dag +m)+\lambda_{a\sigma}'(a^\dag \sigma+a\sigma^\dag)\nonumber\\
&+\xi_d(a^\dag +a)+\xi_p(\sigma^\dag +\sigma).
 \end{align}
Taking dissipations into consideration, the dissipative dynamics of the system can be described by the quantum master equation
\begin{align}\label{eqI03}
\frac{d}{dt}\rho=&-i[H_t,\rho]+\frac{\kappa_a}{2}(2a\rho a^\dag-a^\dag a \rho-\rho a^\dag a)+\frac{\kappa_b}{2}(2b\rho b^\dag-b^\dag b \rho-\rho b^\dag b)\nonumber\\
&+\frac{\kappa_m}{2}(2m\rho m^\dag-m^\dag m \rho-\rho m^\dag m)+\frac{\gamma}{2}(2\sigma\rho \sigma^\dag-\sigma^\dag \sigma \rho-\rho \sigma^\dag \sigma).
 \end{align}
We shift the optical, phonon, and mechanical modes with their mean values $\alpha$, $\beta$, and $\mu$, i.e., $a=\alpha+\delta a$, $b=\beta+\delta b$, and $m=\mu+\delta m$, Eq.\,(\ref{eqI03}) becomes
\begin{align}\label{eqI04}
\frac{d}{dt}\rho=&-i[H_t,\rho]+\frac{\kappa_a}{2}[2(\alpha+\delta a)\rho (\alpha^\ast +\delta a^\dag)-(\alpha^\ast +\delta a^\dag) (\alpha+\delta a) \rho-\rho (\alpha^\ast +\delta a^\dag) (\alpha+\delta a)]\nonumber\\
&+\frac{\kappa_b}{2}[2(\beta+\delta b)\rho (\beta^\ast + \delta b^\dag)-(\beta^\ast + \delta b^\dag) (\beta+\delta b) \rho-\rho  (\beta^\ast + \delta b^\dag) (\beta+\delta b)]\nonumber\\
&+\frac{\kappa_m}{2}[2(\mu+\delta m)\rho (\mu^\ast+\delta m^\dag)- (\mu^\ast+\delta m^\dag) (\mu+\delta m) \rho-\rho (\mu^\ast+\delta m^\dag) (\mu+\delta m)]\nonumber\\
&+\frac{\gamma}{2}[2\sigma\rho \sigma^\dag-\sigma^\dag \sigma \rho-\rho \sigma^\dag \sigma]\nonumber\\
=&-i[H_t,\rho]+\frac{\kappa_a}{2}(\alpha^\ast \delta a \rho-\alpha^\ast \rho \delta a -\alpha \delta a^\dag\rho +\alpha \rho \delta a^\dag)\nonumber\\
&+\frac{\kappa_b}{2}(\beta^\ast \delta b \rho-\beta^\ast \rho \delta b -\beta \delta b^\dag\rho +\beta\rho \delta b^\dag)+\frac{\kappa_m}{2}(\mu^\ast \delta m \rho-\mu^\ast \rho \delta m -\mu \delta m^\dag\rho +\mu \rho \delta m^\dag)\nonumber\\
&+\frac{\kappa_a}{2}(2\delta a\rho \delta a^\dag-\delta a^\dag \delta a \rho-\rho \delta a^\dag \delta a)+\frac{\kappa_b}{2}(2\delta b\rho \delta b^\dag-\delta b^\dag \delta b \rho-\rho \delta b^\dag \delta b)\nonumber\\
&+\frac{\kappa_m}{2}(2\delta m\rho \delta m^\dag-\delta m^\dag \delta m \rho-\rho \delta m^\dag \delta m)+\frac{\gamma}{2}(2\sigma\rho \sigma^\dag-\sigma^\dag \sigma \rho-\rho \sigma^\dag \sigma)\nonumber\\
=&-i[H_1+H_2,\rho]+\frac{\kappa_a}{2}(2\delta a\rho \delta a^\dag-\delta a^\dag \delta a \rho-\rho \delta a^\dag \delta a)+\frac{\kappa_b}{2}(2\delta b\rho \delta b^\dag-\delta b^\dag \delta b \rho-\rho \delta b^\dag \delta b)\nonumber\\
&+\frac{\kappa_m}{2}(2\delta m\rho \delta m^\dag-\delta m^\dag \delta m \rho-\rho \delta m^\dag \delta m)+\frac{\gamma}{2}(2\sigma\rho \sigma^\dag-\sigma^\dag \sigma \rho-\rho \sigma^\dag \sigma),
 \end{align}
where the Hamiltonians $H_1$ and $H_2$ are, respectively, 
\begin{align}\label{eqI05}
H_1=&(\omega_a-\omega_d+\Delta_F)|\alpha|^2+\omega_b|\beta|^2+\omega_m |\mu|^2+\lambda_{ab}'|\alpha|^2(\beta^\ast+\beta)+\lambda_{am}'|\alpha|^2(\mu^\ast +\mu)+\xi_d(\alpha^\ast +\alpha)\nonumber\\
&+[(\omega_a-\omega_d+\Delta_F)\alpha^\ast+\lambda_{ab}' \alpha^\ast(\beta^\ast+\beta)+\lambda_{am}'\alpha^\ast (\mu^\ast+\mu)+\xi_d+\frac{i}{2}\alpha^\ast \kappa_a]\delta a\nonumber\\
&+[(\omega_a-\omega_d+\Delta_F)\alpha+\lambda_{ab}' \alpha(\beta^\ast+\beta)+\lambda_{am}'\alpha (\mu^\ast+\mu)+\xi_d-\frac{i}{2}\alpha \kappa_a]\delta a^\dag\nonumber\\
&+(\omega_b\beta^\ast+\lambda_{ab}'|\alpha|^2+\frac{i}{2}\beta^\ast \kappa_b)\delta b+(\omega_b \beta+\lambda_{ab}'|\alpha|^2-\frac{i}{2}\beta \kappa_b)\delta b^\dag\nonumber\\
&+(\omega_m\mu^\ast+\lambda_{am}'|\alpha|^2+\frac{i}{2}\mu^\ast \kappa_m)\delta m+(\omega_m \mu+\lambda_{am}'|\alpha|^2-\frac{i}{2}\mu \kappa_m)\delta m^\dag,
 \end{align}
and 
\begin{align}\label{eqI06}
H_2=&[\omega_a-\omega_d+\Delta_F+\lambda_{ab}'(\beta^\ast+\beta)+\lambda_{am}'(\mu^\ast+\mu)]\delta a^\dag \delta a+\omega_b \delta b^\dag \delta b+\omega_m \delta m^\dag \delta m+(\omega_{\sigma}-\omega_{d}) \sigma^\dag \sigma\nonumber\\
&+\lambda_{ab}'(\alpha \delta a^\dag+\alpha^\ast \delta a)(\delta b^\dag+\delta b)+\lambda_{am}'(\alpha \delta a^\dag+\alpha^\ast \delta a)(\delta m^\dag+\delta m)+\lambda_{a\sigma}' (\delta a \sigma^\dag+\delta a^\dag \sigma)+\lambda_{a\sigma}'(\alpha\sigma^\dag+\alpha^\ast \sigma)\nonumber\\
&+\lambda_{ab}'\delta a^\dag \delta a(\delta b^\dag +\delta b)+\lambda_{am}'\delta a^\dag \delta a(\delta m^\dag +\delta m)+\xi_p(\sigma^\dag +\sigma ).
 \end{align}
Setting $\lambda_{ab}=\lambda_{ab}'|\alpha|$, $\lambda_{am}=\lambda_{am}'|\alpha|$,  $\lambda_{a\sigma}=\lambda_{a\sigma}'$, and $\xi=\lambda_{a\sigma}'|\alpha|+\xi_p$, the above equation  Eq.\,(\ref{eqI06}) becomes
%
%
\begin{align}\label{eqI08}
H_2=&[\omega_a-\omega_d+\Delta_F+\lambda_{ab}'(\beta^\ast+\beta)+\lambda_{am}'(\mu^\ast+\mu)]\delta a^\dag \delta a+\omega_b \delta b^\dag \delta b+\omega_m \delta m^\dag \delta m+(\omega_{\sigma}-\omega_{d}) \sigma^\dag \sigma\nonumber\\
&+\lambda_{ab}(\delta a^\dag+ \delta a)(\delta b^\dag+\delta b)+\lambda_{am}(\delta a^\dag + \delta a)(\delta m^\dag+\delta m)+\lambda_{a\sigma} (\delta a \sigma^\dag+\delta a^\dag \sigma)+\xi(\sigma^\dag+ \sigma)\nonumber\\
&+\lambda_{ab}'\delta a^\dag \delta a(\delta b^\dag +\delta b)+\lambda_{am}'\delta a^\dag \delta a(\delta m^\dag +\delta m).
 \end{align}
The steady-state mean values $\alpha$, $\beta$, and $\mu$ satisfy 
\begin{align}\label{eqI09}
&(\omega_a-\omega_d+\Delta_F)\alpha+\lambda_{ab}' \alpha(\beta^\ast+\beta)+\lambda_{am}'\alpha (\mu^\ast+\mu)+\xi_d-\frac{i}{2}\alpha \kappa=0,\nonumber\\
&\omega_b \beta+\lambda_{ab}'|\alpha|^2-\frac{i}{2}\beta \kappa_b=0,\nonumber\\
&\omega_m \mu+\lambda_{am}'|\alpha|^2-\frac{i}{2}\mu \kappa_m=0,
 \end{align}
the system dynamics is then decided by
\begin{align}\label{eqI10}
\frac{d}{dt}\rho=&-i[H_2,\rho]+\frac{\kappa_a}{2}(2\delta a\rho \delta a^\dag-\delta a^\dag \delta a \rho-\rho \delta a^\dag \delta a)+\frac{\kappa_b}{2}(2\delta b\rho \delta b^\dag-\delta b^\dag \delta b \rho-\rho \delta b^\dag \delta b)\nonumber\\
&+\frac{\kappa_m}{2}(2\delta m\rho \delta m^\dag-\delta m^\dag \delta m \rho-\rho \delta m^\dag \delta m)+\frac{\gamma}{2}(2\sigma\rho \sigma^\dag-\sigma^\dag \sigma \rho-\rho \sigma^\dag \sigma).
 \end{align}
Under the condition of strong driving at the optical mode, the higher-order terms $\lambda_{ab}'\delta a^\dag \delta a(\delta b^\dag +\delta b)$ and $\lambda_{am}'\delta a^\dag \delta a(\delta m^\dag +\delta m)$ can be neglected safely because of $|\alpha|\gg 1$ and $\lambda_{ab}', \lambda_{am}'\ll \lambda_{ab}, \lambda_{am}$. Then the system Hamiltonian $H_2$ can be further simplified as
\begin{align}\label{eqI11}
H_{\rm eff}=&[\omega_a-\omega_d+\Delta_F+\lambda_{ab}'(\beta^\ast+\beta)+\lambda_{am}'(\mu^\ast+\mu)]\delta a^\dag \delta a+\omega_b \delta b^\dag \delta b+\omega_m \delta m^\dag \delta m+(\omega_{\sigma}-\omega_{d}) \sigma^\dag \sigma\nonumber\\
&+\lambda_{ab}(\delta a^\dag+ \delta a)(\delta b^\dag+\delta b)+\lambda_{am}(\delta a^\dag + \delta a)(\delta m^\dag+\delta m)+\lambda_{a\sigma} (\delta a \sigma^\dag+\delta a^\dag \sigma)+\xi(\sigma^\dag+ \sigma).
 \end{align}
Assuming $\delta a\to a$, $\delta b\to b$, and $\delta m\to m$, the above equation becomes
\begin{align}\label{eqI12}
H=&[\omega_a-\omega_d+\Delta_F+\lambda_{ab}'(\beta^\ast+\beta)+\lambda_{am}'(\mu^\ast+\mu)]a^\dag a+\omega_b b^\dag  b+\omega_m m^\dag m+(\omega_{\sigma}-\omega_{d}) \sigma^\dag \sigma\nonumber\\
&+\lambda_{ab}(a^\dag+ a)(b^\dag+b)+\lambda_{am}(a^\dag + a)(m^\dag+m)+\lambda_{a\sigma} (a \sigma^\dag+a^\dag \sigma)+\xi(\sigma^\dag+ \sigma)\nonumber\\
=&[\omega_a-\omega_d+\Delta_F+\lambda_{ab}'(\beta^\ast+\beta)+\lambda_{am}'(\mu^\ast+\mu)]a^\dag a+\omega_b b^\dag  b+\omega_m m^\dag m\nonumber\\
&+[\omega_{\sigma}-\omega_a+\omega_a-\omega_{d}+\lambda_{ab}'(\beta^\ast+\beta)+\lambda_{am}'(\mu^\ast+\mu)-\lambda_{ab}'(\beta^\ast+\beta)-\lambda_{am}'(\mu^\ast+\mu)] \sigma^\dag \sigma\nonumber\\
&+\lambda_{ab}(a^\dag+ a)(b^\dag+b)+\lambda_{am}(a^\dag + a)(m^\dag+m)+\lambda_{a\sigma} (a \sigma^\dag+a^\dag \sigma)+\xi(\sigma^\dag+ \sigma)\nonumber\\
=&(\Delta_{ad}+\Delta_F)a^\dag a+\omega_b b^\dag  b+\omega_m m^\dag m+(\Delta_{ad}+\Delta_{\sigma a}) \sigma^\dag \sigma\nonumber\\
&+\lambda_{ab}(a^\dag+ a)(b^\dag+b)+\lambda_{am}(a^\dag + a)(m^\dag+m)+\lambda_{a\sigma} (a \sigma^\dag+a^\dag \sigma)+\xi(\sigma^\dag+ \sigma),
 \end{align}
where $\Delta_{ad}=\omega_a-\omega_d+\lambda_{ab}'(\beta^\ast+\beta)+\lambda_{am}'(\mu^\ast+\mu)$ and $\Delta_{\sigma a}=\omega_{\sigma}-\omega_a-\lambda_{ab}'(\beta^\ast+\beta)-\lambda_{am}'(\mu^\ast+\mu)$.

~



 \section{II.\,Derivation of multiquanta resonances conditions}\label{section2}

In the large pumping regime, the strong driving laser can dress the atom, and the system forms new eigenstates that are a quantum superposition of the bare states $\{|e\rangle, |g\rangle\}$. The subsystem Hamiltonian for the strongly driven atom is 
\begin{align}\label{eqII01}
H_{\sigma}=&(\Delta_{\sigma a}+\Delta_{ad})\sigma^\dag \sigma+\xi(\sigma^\dag +\sigma ),
 \end{align}
with the eigenvalues
\begin{align}\label{eqII02}
E_{|\pm\rangle}=\frac{\Delta_{\sigma a}+\Delta_{ad}\pm\sqrt{( \Delta_{\sigma a}+\Delta_{ad} )^2 + 4\xi^2}}{2},
 \end{align}
and corresponding eigenstates
\begin{align}\label{eqII03}
&|+\rangle=c_+|g\rangle+c_-|e\rangle,~~~ |-\rangle=c_-|g\rangle-c_+|e\rangle,
 \end{align}
where
\begin{align}\label{eqII04}
&c_{\pm}=\sqrt{\frac{2\xi^2}{(\Delta_{\sigma a}+\Delta_{ad})^2+4\xi^2\pm(\Delta_{\sigma a}+\Delta_{ad})\sqrt{(\Delta_{\sigma a}+\Delta_{ad})^2+4\xi^2}}}
 \end{align}
 and $c_+^2+c_-^2=1$. Together with the photon, phonon, and magnon modes, and ignoring the influences of the JC interaction and the linear interactions in the optomagnetic system on the energy structure, the eigenvalues of the system Hamiltonian become
\begin{align}\label{eqII05}
E_{|n_a n_b n_m\pm\rangle}=n_a(\Delta_{ad}+\Delta_F)+n_b\omega_b+n_m\omega_m+\frac{\Delta_{\sigma a}+\Delta_{ad}\pm\sqrt{( \Delta_{\sigma a}+\Delta_{ad} )^2 + 4\xi^2}}{2}.
 \end{align}
 In the following, we discuss the resonance conditions considered in the main text. For convenience, we always fix the resonator spinning along the counterclockwise direction. The Fizeau shifts $\Delta_F>0$ and $\Delta_F<0$ correspond to the situations of driving the resonator from the left and right, respectively. 
 
 First, when the spinning resonator is driven from the left, and the total energy of a single clockwise photon and a single magnon matches the transition between the states $|+\rangle$ and $|-\rangle$, i.e.,  $E_{|000+\rangle}=E_{|1 0 1 -\rangle}$, we have
\begin{align}\label{eqII06}
(\Delta_{ad}+|\Delta_F|)+\omega_m-\sqrt{( \Delta_{\sigma a}+\Delta_{ad} )^2 + 4\xi^2}=0,
 \end{align}
 and
\begin{align}\label{eqII07}
\Delta_{ad}=\frac{\Delta_{\sigma a}^2+4\xi^2-(\omega_m+|\Delta_F|)^2}{2(\omega_m+|\Delta_F|-\Delta_{\sigma a})}.
 \end{align}
In other words, these two states $|000+\rangle $ and $|1 01 -\rangle$ are degenerate when the atom is driven at the photon-magnon resonance. When the spinning resonator is driven from the right, and the total energy of a single counterclockwise photon and a single magnon matches with the transition between the states $|+\rangle$ and $|-\rangle$, the photon-magnon resonance condition becomes 
\begin{align}\label{eqII08}
\Delta_{ad}=\frac{\Delta_{\sigma a}^2+4\xi^2-(\omega_m-|\Delta_F|)^2}{2(\omega_m-|\Delta_F|-\Delta_{\sigma a})}.
 \end{align}
The above results correspond to the excitation of the photon-magnon state $|1 01 -\rangle$. The parameter conditions are in agreement with the results shown in the blue area of Fig.1(c) and the orange area of Fig.1(c) in the main text, respectively.

Second, when the spinning resonator is driven from the left, and the total energy of a single clockwise photon and a single phonon matches the transition between the states $|+\rangle$ and $|-\rangle$, i.e.,  $E_{|000+\rangle}=E_{|1 1 0 -\rangle}$ (the states $|000+\rangle $ and $|110 -\rangle$ are degenerate), we have
\begin{align}\label{eqII09}
(\Delta_{ad}+|\Delta_F|)+\omega_b-\sqrt{( \Delta_{\sigma a}+\Delta_{ad} )^2 + 4\xi^2}=0,
 \end{align}
 and
\begin{align}\label{eqII10}
\Delta_{ad}=\frac{\Delta_{\sigma a}^2+4\xi^2-(\omega_b+|\Delta_F|)^2}{2(\omega_b+|\Delta_F|-\Delta_{\sigma a})}.
 \end{align}
 When the spinning resonator is driven from the right, and the total energy of a single counterclockwise photon and a single phonon matches with the transition between the states $|+\rangle$ and $|-\rangle$, the photon-phonon resonance condition becomes 
\begin{align}\label{eqII11}
\Delta_{ad}=\frac{\Delta_{\sigma a}^2+4\xi^2-(\omega_b-|\Delta_F|)^2}{2(\omega_b-|\Delta_F|-\Delta_{\sigma a})}.
 \end{align}
 The above results correspond to the excitation of the photon-phonon state $|1 10 -\rangle$. The parameter conditions are in agreement with those displayed  in the orange area of Fig.1(c) and the pink area of Fig.1(d) in the main text, respectively.

Lastly, because of the system parameters chosen in the main text, the results from the right sides of Eq.(\ref{eqII08}) and Eq.(\ref{eqII10}) are the same, as shown in the orange areas of Figs.1(c) and 1(d). This demonstrates that, in this parameter regime,  the photon-phonon state is resonantly excited when  the resonator is driven from the left, while the photon-magnon state is also resonantly excited when the resonator is driven from the right.

\section{III.\,The optimal local operators in quantum Fisher information}
Multipartite entanglement can be characterized based on quantum Fisher information (QFI) and covariance. For  an arbitrary separable quantum states of $N$ particles, the QFI and covariance must satisfy the inequality\,\cite{Gessner2016PS,Tian2002XS}
 \begin{align}\label{eqIII01}
F_Q[\rho_{\rm sep}, \sum_{j=1}^N A_j]\,\leq 4\sum_{j=1}^N {\rm Var}(A_j)_{\rho_{\rm sep}}\equiv B_n,
 \end{align}
where $A_j$ is a local observable for the $j$th particle,  and ${\rm Var} (A_j)_{\rho}=\langle A_j^2\rangle_{\rho}-\langle A_j\rangle_{\rho}^2$.  The violation of Eq.\,(\ref{eqIII01}) for any choice of $A_j$ is a sufficient criterion for entanglement. However, certain choices of operators $A_j$ may be better suited than others to detect the entanglement of a given state $\rho$. The optimal local operator can be given by  a combination of accessible operators  $A_j=\sum_m c_j^{(m)}A_j ^{(m)}={\bf c}_j \cdot {\bf A}_j$, where  ${\bf A}_j=(A_j^{(1)}, A_j^{(2)}, \dots)^T$ and ${\bf c}_j=(c_j^{(1)}, c_j^{(2)}, \dots)$ are the set of operators for $j$th particle and corresponding coefficients. The full operator is given by $A({\bf c})=\sum_{j=1}^N {\bf c}_j \cdot {\bf A}_j={\bf c}\cdot{\bf A}$, with the  combined vectors ${\bf c}=({\bf c}_1, {\bf c}_2, \dots, {\bf c}_N)$ and the operators ${\bf A}=[{\bf A}_1, \dots, {\bf A}_N]$. 
To detect the  multipartite entanglement in our model,  we choose the family of accessible operators ${\bf A}_j= (\sigma_j^x, \sigma_j^y, \sigma_j^z)^T $ for the atom and ${\bf A}_j= (x_j, p_j, x_j^2, p_j^2, (x_jp_j+p_jx_j)/2)^T $ for the mode $o$ ($o=\{a,b,m\}$),  where $x=o+o^\dag$   and $p=-i(o-o^\dag)$\,\cite{Tian2002XS}. Then the optimal local operators can be given by finding the optimal combined vectors ${\bf c}$. According to Eq.\,(\ref{eqIII01}), $F_Q[\rho, A({\bf c})]-4\sum_{j=1}^N {\rm Var}({\bf c}_j \cdot {\bf A}_j)$ must be nonpositive for arbitrary choices of ${\bf c}$ whenever $\rho$ is separable. To find the optimal ${\bf c}$, we expand the QFI and covariance in their matrix forms. The QFI can be written as  $F_Q[\rho, A({\bf c})]={\bf c}Q_{\rho}^{\mathcal{A}} {\bf c}^T$\,\cite{Hyllus2010GS}, where $Q_{\rho}^{\mathcal{A}}$ is the QFI matrix with  elements 
 \begin{align}\label{eqIII02}
(Q_{\rho}^{\mathcal{A}})_{jj'}^{mm'}=2\sum_{k,k'}\frac{(p_k-p_{k'})^2}{p_k+p_{k'}}\langle \psi_{k}|A_j^{(m)}|\psi_{k'}\rangle \langle \psi_{k'}|A_{j'}^{(m')}|\psi_{k}\rangle
 \end{align}
on the basis of spectral decomposition $\rho=\sum_k p_k |\psi_k\rangle \langle \psi_k|$\,\cite{Braunstein1994Caves}. The covariance can be written as  $\sum_{j=1}^N {\rm Var}({\bf c}_j \cdot {\bf A}_j)={\bf c}\Gamma_{\rho}^{\mathcal{A}} {\bf c}^T$,  where the covariance matrix $\Gamma_{\rho}^{\mathcal{A}}$ has elements $(\Gamma_{\rho}^{\mathcal{A}})_{jj'}^{mm'}={\rm Cov} (A_j^{(m)},A_{j'}^{(m')})_{\rho}$ and $(\Gamma_{\rho}^{\mathcal{A}})_{jj}^{mm'}=0$. Combining the quantum Fisher matrix $Q_{\rho}^{\mathcal{A}}$ with the covariance matrix $\Gamma_{\rho}^{\mathcal{A}}$, the separability criterion reads
 \begin{align}\label{eqIII03}
{\bf c}(Q_{\rho}^{\mathcal{A}} -\Gamma_{\rho}^{\mathcal{A}}){\bf c}^T<0.
 \end{align}
The optimal   ${\bf c}$ can be given by finding the optimal eigenvalue and corresponding eigenstate of the matrix $Q_{\rho}^{\mathcal{A}} -\Gamma_{\rho}^{\mathcal{A}}$. Further, we obtain  the optimal local operators $A_j$ for the entanglement witness.  

A photon-phonon separable state can be described as $\rho_{ab}=\sum_k p_k\rho_a^k\otimes\rho_b^k$ ($\rho_j$ is the reduced density operator).  The right-hand side of of Eq.\,(\ref{eqIII01}) characterizes the bounds of a given state and local operator $A_j$. In our case, we have 
 \begin{align}\label{eqIII04}
&B_1(\rho_{ab})=4 [{\rm Var}(A_a){\rho_a} +{\rm Var}(A_b)_{\rho_b}],\nonumber\\
&B_2(\rho_{ab})=4 [{\rm Var}(A_a+A_b)_{\rho_{ab}} ].
 \end{align}
If the quantity $W^{ab}_1=F_Q[\rho_{ab}, A({\bf c})]-4\sum_{j=a,b} {\rm Var} (A_j)_{\rho_j}=F_Q[\rho_{ab}, A({\bf c})]-B_1(\rho_{ab})>0$, the photon-phonon state is not separable, i.e., there is entanglement between the photon mode and phonon mode. The inequality $F_Q[\rho_{ab}, A({\bf c})]\leq B_2(\rho_{ab})$  is a bound valid for all physical states $\rho_{ab}$.

A photon-phonon-atom fully separable state can be described as $\rho_{ab\sigma}=\sum_{k'} p_{k'}\rho_a^{k'}\otimes\rho_b^{k'}\otimes\rho_{\sigma}^{k'}$.  In this case,  the bounds become
 \begin{align}\label{eqIII05}
&B_1(\rho_{ab\sigma})=4 [{\rm Var}(A_a){\rho_a} +{\rm Var}(A_b)_{\rho_b} +{\rm Var}(A_{\sigma})_{\rho_{\sigma}}],\nonumber\\
&B_2(\rho_{ab\sigma})=4\,{\rm max}\{{\rm Var}(A_a+A_b)_{\rho_{ab}}+{\rm Var}(A_{\sigma})_{\rho_{\sigma}}, {\rm Var}(A_a+A_{\sigma})_{\rho_{a\sigma}}+{\rm Var}(A_{b})_{\rho_{b}}, {\rm Var}(A_b+A_{\sigma})_{\rho_{b\sigma}}+{\rm Var}(A_{a})_{\rho_{a}} \},\nonumber\\
&B_3(\rho_{ab\sigma})=4 [{\rm Var}(A_a+A_b+A_{\sigma})\rho_{ab\sigma}].
 \end{align}
If $W^{ab\sigma}_1=F_Q[\rho_{ab\sigma}, A({\bf c})]-4\sum_{j=a,b,\sigma} {\rm Var} (A_j)_{\rho_j}=F_Q[\rho_{ab\sigma}, A({\bf c})]-B_1(\rho_{ab\sigma})>0$, there is entanglement between the atom, photon mode, and phonon mode. Furthermore, if $W^{ab\sigma}_2=F_Q[\rho_{ab\sigma}, A({\bf c})]-B_2(\rho_{ab\sigma})>0$, the state is fully inseparable, i.e.,  there is a fully inseparable photon-magnon-atom tripartite entanglement\,\cite{Tian2002XS}. The inequality $F_Q[\rho_{ab\sigma}, A({\bf c})]\leq B_3(\rho_{ab\sigma})$  is a bound valid for all physical states $\rho_{ab\sigma}$.
Here it is only necessary to replace mode $b$ with mode $m$ to characterize the entanglement between the magnon mode and others.

 \section{IV.\,Discussion of Experimental feasibility}
Regarding experimental implementations, our proposal could be implemented in a hybrid setup\,\cite{Aspelmeyer2014KM, Chai2022SZ}, where  a coherently driven two-level atom is coupled to a spinning a spinning yttrium iron garnet (YIG) microsphere fixed on a rotating platform\,\cite{Maayani2018DK}. The microsphere supports a mechanical breathing mode and two countercirculating optical whispering gallery modes (WGMs), with the WGM being  strongly driven  by the input light\,\cite{Aspelmeyer2014KM}. Access to the input and output light fields of the microresonator is provided by a tapered optical fiber coupler interfaced with the resonator. The WGMs are modulated by the mechanical breathing mode when the input light pumps at the WGM. Moreover,  because of the spin-orbit coupling of light, the WGMs have a spin along the $z$ direction. By applying a magnetic field parallel to the equator to the microsphere and exciting the magnon mode with an antenna coupling to a microwave field, the WGMs could also be modulated by the dynamic magnetic field via the Faraday effect\,\cite{Chai2022SZ}. Here, scaling down the microsphere size to reduce the mode volumes of both magnon and optical fields, or engineering the microresonator structure such as microrings to increase the overlap between optical and magnon fields, can enhance the coupling between the optical mode and magnon mode. The effective linear coupling strength between the optical and magnon (mechanical) mode that is proportional to the classical cavity amplitude can also be adjusting by the strong driving on the optical mode. 

Moreover, the selection of microresonator is not limited to a microsphere and can also be replaced with a toroidal or bottle microresonator\,\cite{Zhang2016ZZ}. The two-level atom of choice can be a cesium  atom (e.g., the $6 S_{1/2}, F=4 \to 6 P_{3/2}, F'=5$ transition in cesium atom)\,\cite{Aoki2006DW, Dayan2008PA, Alton2011SA}  or rubidium atom (e.g., the $5 S_{1/2}, F=3 \to 5 P_{3/2}, F'=4$ transition in rubidium atom)\,\cite{Junge2013OSV, OShea2013JV, Shomroni2014RL, Will2021MR}.   The atom can be coupled with the microresonator by trapping the atom in the vicinity of the microresonator. However, trapping a single atom in the vicinity of a  microresonator is indeed a daunting task due to several technical and physical complexities. First, precisely positioning and maintaining a single atom near a microresonator requires extremely high stability and accuracy in spatial control, since the optical fields around microresonators can vary significantly over nanometer scales. The use of optical tweezers or other trapping methods to hold a single atom in place involves complex manipulations of light fields. The intensity, phase, and polarization of these light fields must be precisely controlled to create stable trapping potentials without causing excessive heating or ionization of the atom. Second, efficient coupling of the atom to the microresonator's evanescent field requires the atom to be placed within a fraction of the wavelength of the light used, further complicating the spatial control challenge. Lastly, atoms must typically be cooled to very low temperatures to reduce thermal motion, which otherwise could lead to loss of control over the atom's position relative to the microsphere. Any perturbations from the environment (like mechanical vibrations, electromagnetic noise, etc.) can displace the atom from its optimal position, making sustained interaction with the microresonator difficult. 

In recent years, several experiments have successfully achieved coupling between individual cesium atoms (or rubidium atoms) and various types of microresonators, including microspheres, microtoroidal resonators, and bottle microresonators\,\cite{OShea2013JV, Shomroni2014RL, Will2021MR,  Aoki2006DW, Alton2011SA, Dayan2008PA, Junge2013OSV}. Initially, these experiments utilize a magneto-optical trap to trap and cool atomic clouds in a `magneto-optical trap chamber'. Subsequently, the trapping and cooling beams are switched off, allowing the atoms to fall onto the microresonator, thus facilitating the coupling of the individual atoms with the evanescent field of the microresonator’s WGMs. Furthermore, the addition of an optical dipole trap—created by the interference of the incoming laser beam with its reflection from the microresonator surface—is more likely to trap the falling atoms, thereby increasing the interaction time between the atom and the microresonator\,\cite{Will2021MR, Thompson2013TdL}. Based on the experimental groundwork, our model can also employ the techniques described in these studies to achieve the trapping of a single atom and its coupling with a microresonator. In our approach, the atom is initially cooled and trapped within a chamber using a magneto-optical trap, then transported to the vicinity of the microresonator. Here, the microresonator’s WGM evanescent field, which decays exponentially outside the resonator’s surface, interacts with the atom\,\cite{OShea2013JV, Shomroni2014RL, Will2021MR,  Aoki2006DW, Alton2011SA, Dayan2008PA, Junge2013OSV,Thompson2013TdL}. An optical dipole trap may be introduced to enhance the atom's trapping and prolong its interaction time with the microresonator\,\cite{Will2021MR, Thompson2013TdL}. The decay rate of the cesium or rubidium atom  used in trapping near a microresonator has linewidth of tens of ${\rm MHz}$. The coupling strength between the selected atom and the microresonator depends on the properties of the microresonator and the distance between the atom and the microresonator's surface.  Placing the atom closer to the surface of the microresonator\,\cite{Alton2011SA, Will2021MR}, or reducing the size of the microresonator to reduce the mode volume\,\cite{Buck2003Kimble, Spillane2005KV}, can enhance this coupling strength.  H.\,J.\,Kimble and his collaborators, using finite element models, have given the atom-microresonator coupling rate can exceed $2\pi \times700\,{\rm MHz}$ in microresonator  structure given suitable experimental parameters\,\cite{Buck2003Kimble, Spillane2005KV}. In addition to these microresonator-specific techniques, the atom-microresonator coupling can also be enhanced by other physical methods. For instance, both theoretical and experimental evidence suggests that a quantum squeezing method can enhance interactions between quantum systems, even without precise knowledge of the system parameters\,\cite{Lu2015WJ, Qin2018ML, Burd2021SK, Burd2024KS, Qin2024Nori}. Additionally,  the counting of phonons  (magnons) and other types of statistical processing of photons and phonons (magnons)  can be measured by applying an auxiliary system to convert the mechanical (magnon) signals\,\cite{Meenehan2015MM,Bender2019KB}. The multipartite entanglement can be characterized by a class of nonlinear squeezing parameters \,\cite{Tian2002XS,Gessner2019SP}.  For this special design, we theoretically  predicts that the nonreciprocal entangled photon-phonon pairs and photon-magnon pairs with corresponding antibunching $g^{(2)}_{2,ab}=0.29$ and $g^{(2)}_{2,am}=0.49$  could be achieved with the parameters $\omega_b=2\pi\times2\,{\rm GHz}$, $\omega_{m}=2\pi\times2.1\,{\rm GHz}$,  $\Delta_{\sigma a}=-2\pi\times6.2\,{\rm GHz}$,  $|\Delta_{F}|=2\pi\times50\,{\rm MHz}$, $\lambda_{a\sigma}=2\pi\times600\,{\rm MHz}$, $\lambda_{ab}=\lambda_{am}=2\pi\times44\,{\rm MHz}$, $\xi=2\pi\times1.6\,{\rm GHz}$, $\gamma=2\pi\times2\,{\rm MHz}$, and $\kappa=2\pi\times16\,{\rm MHz}$\,\cite{Aspelmeyer2014KM, Chai2022SZ, Zhang2016ZZ, OShea2013JV, Shomroni2014RL, Will2021MR,  Aoki2006DW, Alton2011SA, Dayan2008PA, Junge2013OSV, Buck2003Kimble, Spillane2005KV, Thompson2013TdL}.

\end{document}